\documentclass[12pt,a4paper]{amsart}
\usepackage{amsmath,amstext,amssymb,amsthm,amsfonts}
\newcommand{\nc}{\newcommand}

\nc{\one}{\mbox{\bf 1}}
\nc{\invtensor}{\underset{\leftarrow}{\otimes}}
\nc{\const}{\operatorname{const}}

\nc{\ad}{\operatorname{ad}}
\nc{\tr}{\operatorname{tr}}
\nc{\tp}{\operatorname{top}}
\nc{\rank}{\operatorname{rank}}
\nc{\corank}{\operatorname{corank}}
\nc{\codim}{\operatorname{codim}}
\nc{\sdim}{\operatorname{sdim}}
\nc{\mult}{\operatorname{mult}}

\nc{\spn}{\operatorname{span}}
\nc{\Sym}{\operatorname{Sym}}
\nc{\sym}{\operatorname{sym}}
\nc{\id}{\operatorname{id}}
\nc{\Id}{\operatorname{Id}}
\nc{\Ree}{\operatorname{Re}}

\nc{\htt}{\operatorname{ht}}
\nc{\Ker}{\operatorname{Ker}}
\nc{\rker}{\operatorname{rKer}}
\nc{\im}{\operatorname{Im}}
\nc{\osp}{\mathfrak{osp}}
\nc{\sgn}{\operatorname{sgn}}
\nc{\F}{\operatorname{F}}
\nc{\Mod}{\operatorname{Mod}}
\nc{\Mat}{\operatorname{Mat}}
\nc{\Soc}{\operatorname{Soc}}
\nc{\Inj}{\operatorname{Inj}}
\nc{\Hom}{\operatorname{Hom}}
\nc{\End}{\operatorname{End}}
\nc{\supp}{\operatorname{supp}}
\nc{\Card}{\operatorname{Card}}
\nc{\Ann}{\operatorname{Ann}}
\nc{\Ind}{\operatorname{Ind}}
\nc{\Coind}{\operatorname{Coind}}
\nc{\wt}{\operatorname{wt}}
\nc{\ch}{\operatorname{ch}}
\nc{\Stab}{\operatorname{Stab}}
\nc{\Sch}{{\mathcal S}\mbox{\em ch}}
\nc{\Irr}{\operatorname{Irr}}
\nc{\Spec}{\operatorname{Spec}}
\nc{\Prim}{\operatorname{Prim}}
\nc{\Aut}{\operatorname{Aut}}
\nc{\Fract}{\operatorname{Fract}}
\nc{\gr}{\operatorname{gr}}
\nc{\deff}{\operatorname{def}}
\nc{\HC}{\operatorname{HC}}

\nc{\wdchi}{\widetilde{\chi}}
\nc{\wdH}{\widetilde{H}}
\nc{\wdN}{\widetilde{N}}
\nc{\wdM}{\widetilde{M}}
\nc{\wdO}{\widetilde{O}}
\nc{\wdR}{\widetilde{R}}
\nc{\wdS}{\widetilde{S}}
\nc{\wdV}{\widetilde{V}}

\nc{\wdC}{\widetilde{C}}

\nc{\Ob}{\operatorname{\mathcal Ob}}
\nc{\Dglie}{\operatorname{{\mathcal D}glie}}
\nc{\Fin}{\operatorname{{\mathcal F}in}}

\nc{\Sg}{{\cS(\fg)}}
\nc{\Shg}{{\cS(\fhg)}}
\nc{\Ug}{{\cU(\fg)}}
\nc{\Uhg}{{\cU(\fhg)}}
\nc{\Sh}{{\cS(\fh)}}
\nc{\Uh}{{\cU(\fh)}}
\nc{\Uhh}{{\cU(\fhh)}}
\nc{\Zg}{{{\mathcal{Z}}(\fg)}}

\nc{\Vir}{{\mathcal{V}ir}}

\nc{\tZg}{{\widetilde{\mathcal Z}({\mathfrak g})}}
\nc{\Zk}{{\mathcal Z}({\mathfrak k})}

\nc{\Up}{{\mathcal U}({\mathfrak p})}
\nc{\Ah}{{\mathcal A}({\mathfrak h})}
\nc{\Ag}{{\mathcal A}({\mathfrak g})}
\nc{\Ap}{{\mathcal A}({\mathfrak p})}
\nc{\Zp}{{\mathcal Z}({\mathfrak p})}
\nc{\cZ}{\mathcal Z}
\nc{\cS}{\mathcal S}
\nc{\cP}{\mathcal P}
\nc{\cA}{\mathcal A}
\nc{\cU}{\mathcal U}
\nc{\cH}{\mathcal H}
\nc{\cM}{\mathcal M}
\nc{\cL}{\mathcal L}
\nc{\cF}{\mathcal F}
\nc{\fg}{\mathfrak g}

\nc{\fo}{\mathfrak o}

\nc{\CO}{\mathcal O}
\nc{\Cl}{\mathcal {C}\ell}

\nc{\zq}{\mathpzc q}

\nc{\fl}{\mathfrak l}
\nc{\fn}{\mathfrak n}
\nc{\fm}{\mathfrak m}
\nc{\fp}{\mathfrak p}
\nc{\fh}{\mathfrak h}
\nc{\ft}{\mathfrak t}
\nc{\fk}{\mathfrak k}
\nc{\fb}{\mathfrak b}
\nc{\fs}{\mathfrak s}
\nc{\fB}{\mathfrak B}

\nc{\vareps}{\varepsilon}
\nc{\varesp}{\varepsilon}
\nc{\veps}{\varepsilon}

\nc{\fsl}{\mathfrak{sl}}
\nc{\fgl}{\mathfrak{gl}}
\nc{\fso}{\mathfrak{so}}

\nc{\fpq}{\mathfrak{pq}}
\nc{\fq}{\mathfrak q}
\nc{\fsq}{\mathfrak{sq}}
\nc{\fpsq}{\mathfrak{psq}}

\nc{\fhg}{\hat{\fg}}
\nc{\fhn}{\hat{\fn}}
\nc{\fhh}{\hat{\fh}}
\nc{\fhb}{\hat{\fb}}
\nc{\hrho}{\hat{\rho}}

\nc{\hsl}{\hat{\fsl}}
\nc{\fpo}{\mathfrak{po}}
\nc{\dirlim}{\underset{\rightarrow}{\lim}\,}
\nc{\nen}{\newenvironment}
\nc{\ol}{\overline}
\nc{\ul}{\underline}
\nc{\ra}{\rightarrow}
\nc{\lra}{\longrightarrow}
\nc{\Lra}{\Longrightarrow}

\nc{\Lla}{\Longleftarrow}

\nc{\Llra}{\Longleftrightarrow}

\nc{\thla}{\twoheadleftarrow}

\nc{\hra}{\hookrightarrow}

\nc{\iso}{\overset{\sim}{\lra}}

\nc{\ssubset}{\underset{\not=}{\subset}}

\nc{\vac}{|0;c\rangle}

\nc{\Thm}[1]{Theorem~\ref{#1}}
\nc{\Prop}[1]{Proposition~\ref{#1}}
\nc{\Lem}[1]{Lemma~\ref{#1}}
\nc{\Cor}[1]{Corollary~\ref{#1}}
\nc{\Conj}[1]{Conjecture~\ref{#1}}
\nc{\Claim}[1]{Claim~\ref{#1}}
\nc{\Defn}[1]{Definition~\ref{#1}}
\nc{\Exa}[1]{Example~\ref{#1}}
\nc{\Rem}[1]{Remark~\ref{#1}}
\nc{\Note}[1]{Note~\ref{#1}}
\nc{\Quest}[1]{Question~\ref{#1}}
\nc{\Hyp}[1]{Hypoth\`ese~\ref{#1}}
\nen{thm}[1]{\label{#1}{\bf Theorem.\ } \em}{}
\nen{prop}[1]{\label{#1}{\bf Proposition.\ } \em}{}
\nen{lem}[1]{\label{#1}{\bf Lemma.\ } \em}{}
\nen{cor}[1]{\label{#1}{\bf Corollary.\ } \em}{}
\nen{conj}[1]{\label{#1}{\bf Conjecture.\ } \em}{}

\nen{claim}[1]{\label{#1}{\bf Claim.\ } \em}{}

\nen{defn}[1]{\label{#1}{\bf Definition.\ } }{}
\nen{exa}[1]{\label{#1}{\bf Example.\ } }{}
\nen{rem}[1]{\label{#1}{\em Remark.\ } }{}
\nen{note}[1]{\label{#1}{\em Note.\ } }{}
\nen{exer}[1]{\label{#1}{\em Exercise.\ } }{}
\nen{sket}[1]{\label{#1}{\em Sketch of proof.\ } }{}
\nen{quest}[1]{\label{#1}{\bf Question.\ } \em}{}

\nen{hyp}[1]{\label{#1}{\bf Hypoth\`ese.\ } \em}{}
\begin{document}
\setcounter{section}{-1}

\title[]{Characters of highest weight modules over affine Lie algebras
are meromorphic functions}
\author{Maria Gorelik~$^\dag$}

\address{Dept. of Mathematics, The Weizmann Institute of Science,
Rehovot 76100, Israel}
\email{maria.gorelik@weizmann.ac.il}
\thanks{$^\dag$
Incumbent of the Frances and Max Hersh career development chair.
Supported in part by TMR Grant No. FMRX-CT97-0100.}

\author{Victor Kac~$^\ddag$}

\address{Dept. of Mathematics, 2-178, Massachusetts Institute of Technology,
Cambridge, MA 02139-4307, USA}
\email{kac@math.mit.edu}
\thanks{$^\ddag$ Supported in part by NSF Grant  DMS-0501395.}

\begin{abstract}
We show that the characters of all highest weight modules over 
an affine Lie algebra with the highest weight away from the critical 
hyperplane are meromorphic functions in the positive half of the Cartan 
subalgebra, their singularities being at most simple poles at zeros of 
real roots. We obtain some information about these singularities.
\end{abstract}

\maketitle


\section{Introduction}
\subsubsection{}
Let $\overline{\fg}$ be a simple finite-dimensional Lie algebra over
$\mathbb{C}$, and let $\fg=\overline{\fg}[t,t^{-1}]\oplus\mathbb{C}K
\oplus\mathbb{C}D$ be the associated non-twisted affine Kac-Moody 
algebra~\cite{Kbook}. Recall that the commutation relations
on $\fg$ are:
$$
[at^m,bt^n]=[a,b]t^{m+n}+m\delta_{m,-n}(a|b)K,\ \ [D,at^m]=mat^m,\ \ 
[K,\fg]=0,$$
where $a,b\in\overline{\fg}$,  and $(-,-)$ is a non-degenerate
invariant symmetric bilinear form on $\overline{\fg}$.

Choosing a Cartan subalgebra $\overline{\fh}$ of $\overline{\fg}$, and a 
triangular decomposition $\overline{\fg}=\overline{\fn}_-\oplus
\overline{\fh}\oplus\overline{\fn}_+$, we have the associated Cartan
subalgebra $\fh=\overline{\fh}\oplus\mathbb{C}K\oplus\mathbb{C}D$
and  the triangular decomposition $\fg=\fn_-\oplus\fh\oplus\fn_+$,
where $\fn_{\pm}=\overline{\fn}_{\pm}+\overline{\fg}[t^{\pm 1}]t^{\pm 1}$.
Let $\Delta\supset\Delta_+\supset \Pi=
\{\alpha_0,\alpha_1,\ldots,\alpha_n\}$ be the multiset of
all roots, the multiset of positive roots and the set of simple roots
of $\fg$, respectively.

\subsection{} 
Let $M(\lambda)$ be the Verma module over
$\fg$ with highest weight $\lambda\in\fh^*$. Any quotient 
$V(\lambda)$ of $M(\lambda)$, called a
 highest weight module, has weight space decomposition
$V(\lambda)=\oplus_{\mu\in\fh^*} V_{\mu}(\lambda)$,
where $\dim V_{\mu}(\lambda)<\infty$, hence we can define
its character
$$\ch_{V(\lambda)}(h)=\sum_{\mu} \dim V_{\mu}(\lambda)e^{\mu(h)}.$$

\subsubsection{}\label{Y>}
The character of the Verma module is
\begin{equation}\label{chM}
 \ch_{M(\lambda)}(h)=\frac{e^{\lambda(h)}}{R(h)},\ \ \text{ where }
R(h):=\prod_{\alpha\in\Delta_+} (1-e^{-\alpha(h)}).
\end{equation}
Since $\mult\alpha\leq n=\rank{\fg}$, it is easy to deduce that the series
$\ch_{M(\lambda)}(h)$ converges to a holomorphic function
in the domain
$$Y_{>}:=\{h\in\fh|\ \Ree\alpha_i(h)>0,\ i=0,1,\ldots,n\}.$$
Since $\ch_{V(\lambda)}(h)$ is majorized by $\ch_{M(\lambda)}(h)$,
we deduce that $\ch_{V(\lambda)}(h)$ converges to a holomorphic function
in $Y_{>}$ as well.

\subsection{}
Let $\delta\in\fh^*$, defined by 
$\delta_{\overline{\fh}+\mathbb{C}K}=0,\ \delta(D)=1$, be the minimal
positive imaginary root of $\fg$. Consider the interior of the complexified
Tits cone
$$Y:=\{h\in\fh|\ \Ree\delta(h)>0\}.$$

This is an open domain in $\fh$, which is $W$-invariant,
where $W$ is the Weyl group of $\fg$, and 
$\ol{Y}=\cup_{w\in W} w(\overline{Y}_{>})$~\cite{Kbook}.

It is easy to show that the product 
$R(h)$ converges to a holomorphic function in $Y$, hence
$\ch_{M(\lambda)}(h)$ analytically extends from $Y_{>}$ to
a meromorphic function on the whole domain $Y$. The problem, addressed
in this paper is whether the same holds for $\ch_{V(\lambda)}(h)$.

\subsection{}
Let $\rho\in\fh^*$ be such that
 $(\rho,\alpha_i)=\frac{1}{2}(\alpha_i,\alpha_i)$ for $i=0,1,\ldots,n$.
We prove the following result.
\subsubsection{}
\begin{thm}{}
Let $L(\lambda)$ be the irreducible highest weight module over $\fg$
with highest weight $\lambda$,
such that $(\lambda+\rho)(K)\not=0$. Then $\ch_{L(\lambda)}(h)$
extends from $Y_{>}$ to a meromorphic function in $Y$,
such that its numerator $N_{\lambda}(h):=R(h)\ch_{L(\lambda)}(h)$
is holomorphic in $Y$.
\end{thm}

\subsubsection{}
The proof in the case $(\lambda+\rho)(K)\not\in\mathbb{Q}_{\geq 0}$ 
is very easy (for arbitrary $V(\lambda)$). 
Indeed, it follows easily from~\cite{KK}
(see~\cite{KT}) that, in this case, $N_{\lambda}$ is a finite 
linear combination (over $\mathbb{Z}$) of exponential functions
$e^{\mu}$, hence is a holomorphic function
on the whole space $\fh$. (The difficult problem of computing
the coefficients in $N_{\lambda}$ in this case is solved in~\cite{KT},
but it is not needed here).

\subsubsection{}
In the case $(\lambda+\rho)(K)\in\mathbb{Q}_{>0}$ 
there exists $w\in W(\lambda)$ 
such that $\lambda=x.\lambda'$, where $\lambda'\in Y_{>}$,
$W(\lambda)$ denote the subgroup of $W$, generated by reflections
$s_{\alpha}$ in $\alpha\in\Delta_+$, such that
$(\alpha,\alpha)\not=0$ and $2(\lambda,\alpha)/(\alpha,\alpha)\in\mathbb{Z}$,
and $x.\lambda=x(\lambda+\rho)-\rho$ is the ``shifted action''
of $W$. It follows easily from~\cite{KK} (see~\cite{KT}) that we then have:

\begin{equation}\label{Nla}
N_{\lambda}=\sum_{y\in W(\lambda):\ y\geq x} c_{x,y} e^{y.\lambda},\ \ \ 
c_{x,y}\in\mathbb{Z},
\end{equation}
which is an infinite sum if the group $W(\lambda)$ is infinite.
In order to prove that $N_{\lambda}(h)$ converges
to a holomorphic function in $Y$, we use the explicit formula for the 
$c_{x,y}$, given by the Kazhdan-Lusztig conjecture for $\fg$,
proved in~\cite{KT}:
\begin{equation}\label{cxy}
c_{x,y}=(-1)^{l(y)-l(x)}Q_{x,y}(1),
\end{equation}
where $Q_{x,y}(q)$ are the inverse Kazhdan-Lusztig polynomials
for the Coxeter group $W(\lambda)$.
(The length function $l(x)$ and the Bruhat order $\geq$
in~(\ref{Nla}), (\ref{cxy}) are meant in the group $W(\lambda)$).

\subsection{}
Using the recurrent definition of the polynomials $Q_{x,y}$, we prove
the following estimate (which holds for polynomial growth Weyl groups):
\begin{equation}\label{esti}
|Q_{x,y}(1)|\leq (Cl(y)^{n+1})^{l(y)-l(x)},
\end{equation}
where $C$ is a constant, independent of $x$ and $y$. 
This estimate suffices to prove the theorem.
We also show along the same lines that the theorem holds
for arbitrary highest weight module $V(\lambda)$ with 
$(\lambda+\rho)(K)\not=0$, and arbitrary affine Kac-Moody Lie algebras.
The fact that $\fg$ is affine is used for the obvious, but crucial,
observation that the affine Weyl group has polynomial growth
(since it contains a finitely generated abelian subgroup of finite index).

The theorem probably holds also on the critical hyperplane
$\{\lambda|\ \lambda(K)=-\rho(K)\}$, but it is unclear how to prove it since
already the argument of~\cite{KK} collapses on this hyperplane.

\section{Preliminaries}

\subsection{}\label{not}
Let $A$ be a symmetrizable Cartan matrix, let
$\fg(A)=\fn_-\oplus\fh\oplus\fn$ 
be the corresponding Kac-Moody
algebra over $\mathbb{C}$~\cite{Kbook}, $\Pi$ be the set of simple roots, 
$\Delta_+$ be the {\em multiset} of positive roots,  
and $W$ be the Weyl group. Let $\Delta^{re}$ be the set of real roots
and $\Delta^{im}$ be the multiset of imaginary roots.

\subsubsection{}
Set $Q^+=\{\sum_{\alpha\in\Pi}m_{\alpha}\alpha|\ 
m_{\alpha}\in\mathbb{Z}_{\geq 0}\}$, and
define the standard partial ordering on 
$\fh^*$: $\alpha\geq \beta$ for $\alpha-\beta\in Q^+$.
For $\nu\in Q^+$ denote by $\htt\nu$ the {\em height} of $\nu$:
if $\nu=\sum_{\alpha\in\Pi} m_{\alpha}\alpha$ then
$\htt\nu=\sum_{\alpha\in\Pi} m_{\alpha}$.

\subsubsection{}
Denote by $(-,-)$ a non-degenerate invariant symmetric bilinear form 
on $\fg$. It restricts to a non-degenerate bilinear form
on $\fh$, and the latter induces one on $\fh^*$, also denoted by $(-,-)$.
For $\alpha\in\Delta^{re}$, let
$\alpha^{\vee}\in\fh$ be such that 
$\langle\alpha^{\vee},\mu\rangle=2(\alpha,\mu)/(\alpha,\alpha)$;
then the reflection $s_{\alpha}\in W$ is defined by 
$\mu\mapsto \mu-\langle\alpha^{\vee},\mu\rangle\alpha$, and
$W$ is generated by these reflections.

\subsection{}\label{Cox}
Recall that 
$W$ is a Coxeter group with the canonical system of generators
$\{s_{\alpha}| \alpha\in\Pi\}$ (see~\cite{Kbook}, 3.13).
Denote the unit element
in $W$ by $e$, the Bruhat ordering by $\geq$ ($e$ is the minimal element)
and the length function $l: W\to\mathbb{Z}_{\geq 0}$ by $l(w)$.

Fix $\rho\in\fh^*$ satisfying $\langle \alpha^{\vee},\rho\rangle=1$
for all $ \alpha\in\Pi$ and define the shifted action
of $W$ on $\fh^*$ by $w.\lambda:=w(\lambda+\rho)-\rho$.

\subsubsection{}\label{luara}
For $w\in W$ set
$$S(w):=\Delta_+\cap w^{-1}\Delta_-.$$
For $\alpha\in\Pi$, $w\in W$ one has
$$S(s_{\alpha}w)=\left\{\begin{array}{ll} 
S(w)\cup\{w^{-1}\alpha\}\ \text{ if }
w^{-1}\alpha\in \Delta_+,\\
S(w)\setminus\{-w^{-1}\alpha\}\ \text{ if }
w^{-1}\alpha\in \Delta_-.\end{array}\right.$$
Moreover, $l(s_{\alpha}w)>l(w)$ 
iff $w^{-1}\alpha\in \Delta_+$.
This gives $l(w)=|S(w)|=|w\Delta_+\cap \Delta_-|$
and $l(ws_{\alpha})>l(w)$ 
iff $w\alpha\in \Delta_+$.

We will use the following lemma.
\subsubsection{}
\begin{lem}{lara} (see~\cite{Kbook}, Ex. 3.12)
Let $w=s_{i_1}\ldots s_{i_l}$ be a reduced expression of $w\in W$, where
$s_{i}$ denote the  reflection with respect to the simple
root $\alpha_{i}$.
Set $w^{(0)}=e,\ \ w^{(j)}=s_{i_1}\ldots s_{i_j}$
and  $\beta_j:=w^{(j)}\alpha_{i_{j+1}}$ for $j=0,\ldots,l-1$.
Then $\beta_j$ are pairwise distinct, $\{\beta_j\}_{j=0}^{l-1}=S(w^{-1})$ 
and for any  $\lambda\in\fh^*$
$$\lambda-w\lambda=\sum_{j=0}^{l-1} 
\langle\alpha_{i_{j+1}}^{\vee},\lambda\rangle\beta_j.$$
\end{lem}
\begin{proof}
 The fact that $\{\beta_j\}_{j=0}^{l-1}=S(w^{-1})$ follows from
the above description of $S(s_{\alpha}w)$; since
$|S(w^{-1})|=l(w^{-1})=l$ the elements $\beta_j$ are pairwise distinct.

One has
$$\lambda-w\lambda=w^{(l-1)}(\lambda-s_{i_l}\lambda)+
(\lambda-w^{(l-1)}\lambda)=
\langle\alpha_{i_{l}}^{\vee},\lambda\rangle w^{(l-1)}\alpha_{i_{l}}
+(\lambda-w^{(l-1)}\lambda)$$
so the assertion follows by induction on $l=l(w)$.
\end{proof}

\subsection{}
In this text $\fg$ will denote an (arbitrary) affine Lie algebra,
unless otherwise stated. Let $\delta$
be the minimal positive imaginary root. We fix
the form $(-,-)$ to be positive semidefinite  on 
$\fh^*_{\mathbb{R}}=\sum_{\alpha\in\Pi} \mathbb{R}\alpha$;
its kernel is $\mathbb{R}\delta$.

\subsubsection{}\label{cl}
Set  $E:=\mathbb{R}\Delta^{re}$, $E_{cl}=E/\mathbb{R}\delta$,
and let $cl:E\to E_{cl}$ denote the projection. The form $(-,-)$
on $\fh^*$ induces a symmetric bilinear form on $E_{cl}$ which
is positive definite.
Set $\Delta_{cl}=cl(\Delta^{re})$; this is a finite root system in $E_{cl}$
(not necessarily reduced).

\subsection{Groups of  polynomial growth}
Let $G$ be a finitely generated group and $B$ be 
its finite system of generators. 
For $g\in G$ denote by $l_B(g)$ the ``length''
of $g$, i.e. the smallest natural number $k$ such that $g=e_1e_2\ldots e_k$,
where $e_i\in B\cup B^{-1}$. For $r\in \mathbb{R}_{>0}$ set
$$c_{G,B}(r):=|\{ g\in G|\ l(g)\leq r\}|.$$
If $B'$ is another finite system of generators then for some 
$a,b\in\mathbb{Z}_{>0}$ one has
$$c_{G,B}(ar)\leq c_{G,B'}(r)\leq c _{G,B}(br)\ \  \text{ for all } r.$$
A finitely generated group is called a group
of a {\em polynomial growth} if there exist $C>0, m\geq 0$
such that $c_{G,B}(r)\leq cr^m$ for any $r\geq 1$.
For instance, a finitely generated abelian group is
of  polynomial growth (if the rank is $m$ then 
$c_{G,B}(r)\leq (r+m)^m<Cr^m$ for $r\geq 1$).

Let $G$ be a group and
$G_1$ be its finitely generated subgroup of finite index. Then $G$ is also
finitely generated. Let $B$ (resp., $B_1$) be a  
finite system of generators of $G$ (resp., $G_1$).
Then there exist $a,b,C>0$ such that
$$c_{G_1,B_1}(ar)\leq c_{G,B}(r)\leq Cc _{G_1,B_1}(br)
\ \  \text{ for all } r.$$
As a result, $G$ is of polynomial growth  iff $G_1$ is of 
polynomial growth.

\section{Coxeter subgroups of $W$}
\subsection{}\label{subs}
Following~\cite{KT}, \cite{MP} we call a subset $\Delta_1$ of $\Delta^{re}$
a {\em subsystem} of $\Delta^{re}$ if $s_{\alpha}\beta\in\Delta_1$ for
any $\alpha,\beta\in\Delta_1$. For a subsystem $\Delta_1$ of $\Delta^{re}$
we set
$$\begin{array}{ll}
\Delta_{1,\pm}:=\Delta_{\pm}\cap\Delta_1,\ & \ 
\Pi_1:=\{\alpha\in \Delta_{1,+}| \ 
s_{\alpha}(\Delta_{1,+}\setminus\{\alpha\})\subset \Delta_{1,+}\},\\
W_1:=\langle s_{\alpha}| \alpha\in \Pi_1\rangle,\ &\ 
S_1:=\{s_{\alpha}| \alpha\in \Pi_1\}.
\end{array}$$
Notice that  $\Delta_{1,+}=-\Delta_{1,-}$ since for $\alpha\in\Delta_1$
the element $-\alpha=s_{\alpha}\alpha$ lies in $\Delta_1$.

\subsection{}
\begin{lem}{pim}
(i) For $\alpha,\beta\in\Pi_1$ one has 
$\langle\alpha^{\vee},\beta\rangle\in\mathbb{Z}_{\leq 0}$ if 
$\alpha\not=\beta$.

(ii) $|\Pi_1|\leq |\Delta_{cl}|$ (see~\ref{cl} for notation).
\end{lem}
\begin{proof}
 Take $\alpha\not=\beta$ and assume that $(\alpha,\beta)>0$. 
Clearly, $\langle\alpha^{\vee},\beta\rangle\in\mathbb{Z}$.
Since $(\alpha,\beta)>0$ and $s_{\alpha}\beta\in\Delta_+$ 
we get $\beta\geq \alpha$; similarly, 
$\alpha\geq \beta$, which is a contradiction. Hence $(\alpha,\beta)<0$
so $\langle\alpha^{\vee},\beta\rangle\in\mathbb{Z}_{\leq 0}$ 
as required for (i).
For (ii) notice that $(\alpha,\beta)<0$ 
implies that $\alpha-\beta$ is not proportional to $\delta$
so $cl(\alpha)\not=cl(\beta)$ for distinct $\alpha,\beta\in\Pi_1$.
\end{proof}

\subsection{}\label{secXi}
In the light of~\Lem{pim}
the matrix 
$A:=\left(\langle\alpha^{\vee},\beta\rangle\right)_{\alpha,\beta\in\Pi_1}$
is a generalized Cartan  matrix. Let $X$ be
the Dynkin diagram constructed for $\Pi_1$ and $A$: 
the elements of
$\Pi_1$ are nodes in $X$ and the edges (and  the arrows) are determined
in the usual way by $A$. Note that the elements of $\Pi_1$ are not 
necessarily linearly independent.

Let $X_i, i=1,\ldots,k,$ be the connected components of $X$.
We write $\alpha\in X_i$ if $\alpha$ is a node of $X_i$.
The following result goes back to Dynkin
(see also~\cite{B}). 

\subsubsection{}
\begin{prop}{propXi}
(i) For each $i=1,\ldots,k$ the roots from $X_i$
 are linearly independent.

(ii) Any linear dependence of the elements of $\Pi_1$ is of the form
$\sum_{i=1}^k\sum_{\alpha\in X_i} m_{\alpha}\alpha=0$, where
$\sum_{\alpha\in X_i} m_{\alpha}\alpha\in\mathbb{C}\delta$
for each $i$.

(iii) The Dynkin diagram for $\Pi_1$ is a disjoint union of Dynkin 
diagrams of affine and finite types. 
\end{prop}
\begin{proof}
Any linear dependence $\sum_{\alpha\in \Pi_1} m_{\alpha}\alpha=0$
with $m_{\alpha}\in\mathbb{R}$ 
can be rewritten in the form 
$\sum_{\alpha\in S} m_{\alpha}\alpha=\sum_{\beta\in S'} k_{\beta}\beta$,
where $S,S'$ are non-empty subsets of $\Pi_1$ with empty intersection
$S\cap S'=\emptyset$ and 
all coefficients are positive: $m_{\alpha},k_{\beta}>0$.
Put $\gamma:=\sum_{\alpha\in S} m_{\alpha}\alpha$. Then
$$(\gamma,\gamma)=
\sum_{\alpha\in S,\beta\in S'}m_{\alpha}k_{\beta}(\alpha,\beta)\leq 0.$$
Since $(-,-)$ is semidefinite on $\fh^*_{\mathbb{R}}$ and 
the isotropic vectors are proportional to $\delta$, we get 
that $\gamma\in\mathbb{R}\delta$ and that
$(\alpha,\beta)=0$ for all $\alpha\in S,\beta\in S'$.

Let us show that for any $i$ one has either $X_i\cap S=\emptyset$
or $X_i\subset S$. Indeed, assume that $X_i\cap S\not=\emptyset$ and
$X_i\not\subset S$. Take $\alpha'\in
X_i\setminus S$ which is connected to a node in $X_i\cap S$.
Then 
$$(\gamma,\alpha')=\sum_{\alpha\in S\cap X_i}m_{\alpha}(\alpha,\alpha')<0$$
since $(\alpha,\alpha')\leq 0$ for all $\alpha$ and 
$(\alpha,\alpha')<0$ if $\alpha'$ is connected to $\alpha$.
However $\gamma\in\mathbb{R}\delta$ so
$(\gamma,\alpha')=0$, which is a contradiction.

By above, for any $i$ one has either $X_i\cap S=\emptyset$
or $X_i\subset S$. The similar fact holds for $S'$. Since both $S,S'$
are non-empty, their union does not lie in $X_i$:
$S\cup S'\not\subset X_i$. Hence the roots of $X_i$ are linearly independent
as required for (i). For (ii) write
$\gamma=\sum_{i=1}^k \gamma_i$, where $\gamma_i:=\sum_{\alpha\in X_i}
m_{\alpha}\alpha$ and notice that
$$0=(\gamma,\gamma)=\sum_{i=1}^k (\gamma_i,\gamma_i).$$
Hence $(\gamma_i,\gamma_i)=0$ so $\gamma_i\in\mathbb{R}\delta$
for any $i$. This gives (ii).

For (iii) fix $i$ and let $A_i$ be the corresponding submatrix of $A$:
$A_i:=\left(\langle
\alpha^{\vee},\beta\rangle\right)_{\alpha,\beta\in X_i}$.
Since the nodes of $X_i$ are linearly independent, $X_i$
is a standard Dynkin diagram. The matrix $A_i$ is symmetrizable
and the corresponding bilinear form on $\sum_{\alpha\in X_i}\mathbb{R}\alpha$
is the restriction of $(-,-)$. Therefore
this form is either positive definite or positive semi-definite.
Hence $X_i$ is of either finite type or affine type.
\end{proof}

\subsubsection{}
\begin{cor}{Waf}
(i) The group $W_1$, generated by $S_1$, is a direct
product of Coxeter groups  of finite or affine types corresponding
to the connected components of $X$. 

(ii) The group $W_1$ is the Weyl group 
corresponding to the Dynkin diagram $X$: it
is a Coxeter group with the canonical generator system
$S_1$, and its length function $l_1: W_1\to\mathbb{Z}_{\geq 0}$
is given by $l_1(w)=|w\Delta_{1,+}\cap \Delta_{1,-}|$.
\end{cor}
\begin{proof}
Let $W'_i$ be the Coxeter group corresponding to the Dynkin diagram $X_i$
and $\Delta(X_i)\subset \Delta_1$ be the root system of $X_i$. 
Since the Dynkin diagram of $\Pi_1$ is the disjoint union of $X_i$s,
there exists a homomorphism 
$\phi: W_1'\times W_2'\times\ldots\times W'_k\to W_1$. 
Take $w_1'w_2'\ldots w'_k\in\Ker\phi$, where
$w'_i\in W'_i$.
Assume that $w'_i\not=e$. Then there exists $\alpha\in X_i$ such that
$l_i(s_{\alpha}(w'_i)^{-1})<l_i((w'_i)^{-1})$, where $l_i$ stands
for the length function $l_i: W_i'\to\mathbb{Z}_{\geq 0}$.
By~\ref{Cox} $w'_i\alpha\in\Delta_-\cap \Delta(X_i)$.
The elements $w'_i\alpha,\alpha$ lie in $\Delta(X_i)$ 
so they are stable with respect to the action of $W_j'$ for $j\not=i$.
Thus $(w_1'w_2'\ldots w'_k)\alpha=w'_i\alpha\in\Delta_-$
so $(w_1'w_2'\ldots w'_k)\alpha\not=\alpha$ which contradicts
to the assumption  $w_1'w_2'\ldots w'_k\in\Ker\phi$. Hence
$\Ker\phi=e$ and this proves (i); (ii) follows from (i).
\end{proof}

\subsection{}
\begin{rem}{remKT}
By~\cite{KT15}, 2.2.7, for any root system $\Delta$
one has $\Delta_{1,+}\subset \sum_{\alpha\in\Pi_1}\mathbb{Z}_{\geq 0}\alpha$,
$\Delta_1=W_1\Pi_1$, and so
$W_1=\langle s_{\alpha}| \alpha\in\Delta_1 \rangle$.
\end{rem}

\section{Kazhdan-Lusztig polynomials}
We recall the construction
of  Kazhdan-Lusztig polynomials introduced in~\cite{KL}.
In this section $W$ is a Coxeter group, isomorphic
to the Weyl group of a Kac-Moody algebra.

\subsection{Polynomials $P_{x,y}(q)$}
The Kazhdan-Lusztig polynomials $P_{x,y}(q)\in\mathbb{Z}[q]$ ($x,y\in W$)
can be described recursively in the following way
(see~\cite{KL}, 2.2 c):
 
(a) $P_{x,y}=0$ if $x\not\leq y$; $P_{x,x}=1$;

(b) if $x<y$ and $s$ is a simple reflection such that $sy<y$ then 
$$P_{x,y}=q^{1-c}P_{sx,sy}+q^cP_{x,sy}-\sum_{z:\ x\leq z<sy,\ sz<z} 
q^{\frac{l(y)-l(z)}{2}}\mu(z,sy)P_{x,z},$$
where $c=1$ if $sx<x$, $c=0$ if $sx>x$ and
$\mu(z,w)$ is the coefficient of
$\frac{l(w)-l(z)-1}{2}$th power of $q$
 in $P_{z,w}$ (defined to be $0$ if
$l(ys)-l(z)$ is even). Notice that the degree of 
$P_{x,y}$ is not greater than $\frac{l(y)-l(x)-1}{2}$.

Due to non-negativity of coefficients of $P_{x,y}(q)$~\cite{KL},
\cite{KL2},\cite{Hd}, the
 above recursive  description of $P_{x,y}$ gives 
$P_{x,y}(1)\leq P_{sx,sy}(1)+P_{x,sy}(1)$ if $sy<y$, hence

\begin{equation}\label{Pestim}
P_{x,y}(1)\leq l(y)^{l(y)-l(x)}.
\end{equation}

\subsection{Polynomials $Q_{x,y}(q)$}
The inverse Kazhdan-Lusztig polynomials $Q_{x,y}(q)\in\mathbb{Z}[q]$ 
($x,y\in W$) are defined by the property
$$\sum_w (-1)^{l(w)-l(x)} Q_{x,w} P_{w,y}=\delta_{x,y}.$$
One has $Q_{x,y}=0$ for $y\not\geq x$, $Q_{x,x}=1$, and for $y>x$ 
the degree of $Q_{x,y}$ is not greater than $\frac{l(y)-l(x)-1}{2}$.

\subsubsection{}
For any fixed $z\in W$ the matrix $(P_{w,y})_{w,y\leq z}$
is a square matrix of  finite size; by above, this matrix is inverse
to the matrix $((-1)^{l(w)-l(x)}Q_{x,w})_{x,w\leq z}$. Therefore
$$\sum_{w: x\leq w\leq y} 
(-1)^{l(y)-l(w)} P_{x,w} Q_{w,y}=\delta_{x,y}.$$
Using that $P_{x,x}=1$, we express $Q_{x,y}$ via the rest of summands to obtain
\begin{equation}\label{Qcon}
Q_{x,y}=\sum_{w: x<w\leq y}  
(-1)^{l(w)-l(x)+1}P_{x,w}Q_{w,y}\ \text{ for }\ y>x.
\end{equation}

\subsubsection{}
\begin{lem}{Qestim}
Assume that $W$ has  polynomial growth, i.e.
for some constants 
$C>0,\ n\geq 0$ one has
$$ |\{w|\ l(w)\leq k\}|< Ck^{n},\ \ \text{ for all 
positive integers }\ k.$$
Put $N:=\max (n+1;2)$.
Then for all $x,y\in W$  one has 
$$|Q_{x,y}(1)|\leq (C l(y)^N)^{l(y)-l(x)}.$$
\end{lem}
{\em Proof.}
The proof is by induction on $l:=l(y)-l(x)$. For $l\leq 2$ 
one has $Q_{x,y}=0$ or $1$ so the assertion holds. Assume that $l\geq 3$.
From~(\ref{Qcon}) one sees that
$$|Q_{x,y}(1)|\leq Cl(y)^{n}\max_{w: x<w\leq y}|P_{x,w}(1) Q_{w,y}(1)|.$$
Set $a:=l(x), b:=l(y), t:=l(w)$ and note that $t\in [a+1,b]$.
By~(\ref{Pestim}) one has $P_{x,w}(1)\leq t^{t-a}$. 
The induction hypothesis gives
$$|Q_{x,y}(1)|\leq C^{b-a}b^{n}
\max_{t\in [a+1,b]}\bigl(b^{N(b-t)}t^{t-a}\bigr).$$
Set $g(t):=b^{N(b-t)}t^{t-a}$. Then
$$g'(t)=g(t)(-N\ln b+\ln t+1-a/t).$$
For $t\in [a+1,b]$ one has
$-N\ln b+\ln t+1-a/t<-(N-1)\ln b+1<0$ since $b=l(y)\geq 3$ and $N>1$.
Hence $g'(t)<0$ for $t\in [a+1,b]$ and therefore
$$\max_{t\in [a+1,b]} g(t)=g(a+1)=b^{N(b-a-1)}(a+1)\leq 
b^{N(b-a-1)+1}.$$
Thus
$$|Q_{x,y}(1)|\leq C^{b-a}b^{n+N(b-a-1)+1}=C^{b-a}b^{N(b-a)}. \qed $$

\section{Characters of irreducible modules with non-critical highest
weights}
In this section we recall the results of M.~Kashiwara and T.~Tanisaki
\cite{KT}. Let $\fg$ be an affine Lie algebra and $\lambda\in\fh^*$
be a non-critical weight, i.e. $(\delta, \lambda+\rho)\not=0$.

\subsection{Notations}\label{notKT}
For $\lambda\in\fh^*$ set
denote by $\Delta(\lambda)$ the set of real
roots satisfying $\langle\alpha^{\vee},\lambda+\rho\rangle\in\mathbb{Z}$
and by $\Delta_0(\lambda)$ the set of real
roots satisfying $\langle\alpha^{\vee},\lambda+\rho\rangle=0$.
Clearly, both $\Delta(\lambda)$ and $\Delta_0(\lambda)$ 
are root subsystems in a sense of~\ref{subs}. 
Denote by $W(\lambda)$ (resp., $W_0(\lambda)$) the subgroup
of $W$ generated by $\{s_{\alpha}| \alpha\in \Delta(\lambda)\}$
(resp.,  by $\{s_{\alpha}| \alpha\in \Delta_0(\lambda)\}$).

Since 
$(\delta, \lambda+\rho)\not=0$ one has $cl(\alpha)\not=cl(\beta)$
for $\alpha,\beta\in \Delta_{+,0}(\lambda)$ (see~\ref{cl} for notation).
Therefore $\Delta_{+,0}(\lambda)$ is finite, so  
$W_0(\lambda)$ is a finite Coxeter group.

\subsubsection{}
By~\Rem{remKT}, $W(\lambda)$ 
is a Coxeter group with the canonical generator system
$\{s_{\alpha}| \alpha\in \Pi(\lambda)\}$, where 
$\Pi(\lambda)$ for $\Delta(\lambda)$ is defined in~\ref{subs}.

Denote the Bruhat ordering  of $W(\lambda)$ by $\geq_{\lambda}$ and
and  the length function by $l_{\lambda}: W(\lambda)\to\mathbb{Z}_{\geq 0}$.
For $x,y\in W(\lambda)$ denote by $P^{\lambda}_{x,y}(q)\in\mathbb{Z}[q]$
(resp.,  by $Q^{\lambda}_{x,y}(q)\in\mathbb{Z}[q]$) 
the corresponding Kazhdan-Lusztig polynomial (resp., inverse 
Kazhdan-Lusztig polynomial).

\subsection{}\label{C+-}
Let $\mathcal{C}$ be the set of non-critical weights, i.e.
$$\mathcal{C}:=\{\lambda\in\fh^*|\ (\delta,\lambda+\rho)\not=0\}.$$
Let
$$\begin{array}{l}
\mathcal{C}^+:=\{\lambda\in\mathcal{C}|\ 
\langle\alpha^{\vee},\lambda+\rho\rangle\geq 0\text{ for any } 
\alpha\in \Delta(\lambda)\cap \Delta_+\},\\
\mathcal{C}^-:=\{\lambda\in\mathcal{C}|\ 
\langle\alpha^{\vee},\lambda+\rho\rangle\leq 0\text{ for any } 
\alpha\in \Delta(\lambda)\cap\Delta_+\}.
\end{array}$$

\subsection{}\label{alt}
Observe that $\Delta(\lambda)=\emptyset$ forces
$\Delta(w.\lambda)=\emptyset$ for any $w\in W$. 
Therefore if $\Delta(\lambda)=\emptyset$, then $M(w.\lambda)=L(w.\lambda)$
for any $w\in W$. 

Let $\lambda\in\mathcal{C}$ be such that $\Delta(\lambda)\not=\emptyset$.
In the light of~Lemma 2.10 of~\cite{KT} there are the following cases:

(a) If $(\delta,\lambda+\rho)\not\in\mathbb{Q}$, then 
$|(W(\lambda).\lambda)\cap \mathcal{C}^+|=
|(W(\lambda).\lambda)\cap \mathcal{C}^-|=1$.

(b) If $(\delta,\lambda+\rho)\in\mathbb{Q}_{>0}$, then 
$|(W(\lambda).\lambda)\cap \mathcal{C}^+|=1$  and
$|(W(\lambda).\lambda)\cap \mathcal{C}^-|=0$.

(c) If $(\delta,\lambda+\rho)\in\mathbb{Q}_{<0}$, then 
$|(W(\lambda).\lambda)\cap \mathcal{C}^-|=1$ and
$|(W(\lambda).\lambda)\cap \mathcal{C}^+|=0$.

In particular, if $M(\lambda')$ is not simple then there exists 
$\lambda\in \mathcal{C}^+\cup\mathcal{C}^-$ and $w\in W(\lambda)$
such that $\lambda'=w.\lambda$.

\subsubsection{}
\begin{lem}{sl}
Let $\lambda\in \mathcal{C}^+\cup\mathcal{C}^-$.

(i) The group $W_0(\lambda)$ coincides with the stabilizer of
$\lambda$ in $W(\lambda)$ and is generated
by the set $\{s_{\alpha}| \alpha\in\Pi_0(\lambda)\}$,
where 
$\Pi_0(\lambda)=\Pi(\lambda)\cap \{\alpha| \langle \alpha^{\vee},
\lambda+\rho\rangle=0\}$.

(ii) For any $w\in W(\lambda)$ the coset $wW_0(\lambda)$
contains a unique minimal element and a unique maximal element.
\end{lem}
\begin{proof}
(i) Fix $\lambda\in \mathcal{C}^+$.
Take $w\in W(\lambda)$ and let 
$w=s_{i_1}\ldots s_{i_l}$ be a reduced expression of $w\in W$, where
each $s_{i_j}$ is a simple reflection with respect to 
$\alpha_{i_j}\in\Pi(\lambda)$.
By~\Lem{lara},
$$\lambda-w.\lambda=(\lambda+\rho)-w(\lambda+\rho)=\sum_{j=0}^{l-1} 
\langle\alpha_{i_{j+1}}^{\vee},\lambda+\rho\rangle \beta_j$$
for some $\beta_j\in\Delta_+\cap\Delta(\lambda)$.
The condition $\lambda\in \mathcal{C}^+$
ensures that the coefficient of $\beta_j$ is non-negative for any $j$.
If $\lambda-w.\lambda=0$ this implies
$\langle\alpha_{i_{j+1}}^{\vee},\lambda+\rho\rangle=0$
for any $j$. Therefore the stabilizer of $\lambda$ in $W(\lambda)$
is generated by simple reflections of $W(\lambda)$ and
$W_0(\lambda)$ coincides with this stabilizer. The proof
for $\lambda\in \mathcal{C}^-$ is completely similar.

(ii) Let $x\in wW_0(\lambda)$ be a minimal element.
Let us show that for any $y\in W_0(\lambda)$ one has $l(xy)=l(x)+l(y)$.
The proof is by induction on $l(y)$. Assume that $l(xy)=l(x)+l(y)$
and that  $\alpha\in\Pi_0(\lambda)$ is such that $ys_{\alpha}>y$.
By~\ref{luara}, $ys_{\alpha}>y$ gives  $y(\alpha)\in\Delta_{0,+}(\lambda)$;
from (i) one sees that $y(\alpha)$ is a non-negative
linear combination of elements of $\Pi_0(\lambda)$.
The minimality of $x$ implies $x(\beta)\in\Delta_+$ 
for any $\beta\in\Pi_0(\lambda)$. Therefore $xy(\alpha)\in\Delta_+$
so $l(xys_{\alpha})=l(xy)+1$ by~\ref{luara}. Hence $l(xy)=l(x)+l(y)$
for any $y\in W_0(\lambda)$, and (ii) follows from the finiteness of
$W_0(\lambda)$.
\end{proof}

\subsubsection{}\label{sl1}
Take $\lambda'\in \mathcal{C}$ such that $M(\lambda')$ is not simple
and choose $\lambda\in \mathcal{C}^+\cup\mathcal{C}^-$ such that 
$\lambda'=w.\lambda$ for some $w\in W(\lambda)$.
By~\Lem{sl}, $wW_0(\lambda)=\{x\in W(\lambda)|\ x.\lambda=\lambda'\}$
and this set contains a unique minimal element and a unique maximal element,
which we denote respectively by
$w_s(\lambda';\lambda)$ and  $w_l(\lambda';\lambda)$.

\subsubsection{}
\begin{thm}{thm11} (\cite{KT}, Thm. 1.1).
Take $\lambda'\in \mathcal{C}$ such that $M(\lambda')$ is not simple.
Take $\lambda\in \mathcal{C}^+\cup\mathcal{C}^-$ such that 
$\lambda'=w.\lambda$ for some $w\in W(\lambda)$ and define
$w_s(\lambda';\lambda)$, $w_l(\lambda';\lambda)$ as in~\ref{sl1}.

(i) If $\lambda\in\mathcal{C}^+$ set $x:=w_l(\lambda';\lambda)$
and observe that $\lambda'=x.\lambda$. One has

$$\ch_{L(x.\lambda)}=\sum_{y\in W(\lambda): y\geq_{\lambda} x}
(-1)^{l_{\lambda}(y)-l_{\lambda}(x)}Q^{\lambda}_{x,y}(1)\ch_{M(y.\lambda)}.$$

(ii) If $\lambda\in\mathcal{C}^-$ set $z:=w_s(\lambda';\lambda)$
and observe that $\lambda'=z.\lambda$. One has

$$\ch_{L(z.\lambda)}=\sum_{y\in W(\lambda): y\leq_{\lambda} z}
(-1)^{l_{\lambda}(z)-l_{\lambda}(y)}P^{\lambda}_{y,z}(1)\ch_{M(y.\lambda)}.$$
\end{thm}

\section{Proof of~\Thm{propmain}}
\subsection{}
\begin{thm}{propmain}
Let $V(\Lambda)$ be a highest weight module with highest weight
$\Lambda$ over an affine Lie algebra, and assume that 
$(\Lambda+\rho,\delta)\not=0$.
Then the function
$\ch_{V(\Lambda)}(h)$ is meromorphic in the domain
$Y:=\{h\in\fh|\ \Ree \langle h,\delta\rangle >0\}$.
\end{thm}

\subsubsection{}
\begin{rem}{}
In fact, we will show that $R(h)\ch_{V(\Lambda)}(h)$ 
is holomorphic in $Y$, where
$R$ is the Weyl denominator, which is holomorphic in $Y$ as well
(see~\ref{Rr}).
\end{rem}

\subsection{}\label{Xk}
Write $\Pi=\{\alpha_0,\alpha_1,\ldots,\alpha_n\}$ and 
define a grading on $Q^+$ by setting $\deg\alpha_0=1,\ \deg\alpha_i=1$
for $i=1,\ldots,n$. Consider the induced grading on $\Delta_+$:
$\Delta_+=\cup_{j\geq 0} \Delta_{+,j}$. 
From the structure theory of affine Lie 
algebras (\cite{Kbook}, Ch. VI) one knows that 
the multiset $\Delta$ contains at most $n:=\dim\fh-2$ copies of each
imaginary root which is of the form $k\delta$,
and that all real roots have multiplicity one and are of the form
$k\delta+\alpha$ for $\alpha\in\Delta_{cl}$; in particular,
\begin{equation}\label{Deltaj}
\forall j\ \ \ |\Delta_{+,j}|\leq N, \ \text{ where }
N:=|\Delta_{cl}|+\dim\fh-2.\end{equation} 

\subsection{}\label{Rr}
Consider the infinite product
$$R(h):=\displaystyle\prod_{\alpha\in\Delta_+}
(1-e^{-\alpha(h)}), \ h\in Y.$$
It is well-known  that $R(h)$ converges to a 
holomorphic function in the domain $Y$.
Indeed, using the Weierstrass criterion, it is enough to show that 
$$
\sum_{\alpha\in\Delta_+} |e^{-\langle h,\alpha\rangle}|<\infty \text{ if }
h\in Y.$$
From~\ref{Xk} one sees that
$$\sum_{\alpha\in\Delta_+} 
|e^{-\langle h,\alpha\rangle}|<\bigl(\sum_{\alpha\in\Delta_{cl}}
|e^{-\langle h,\alpha\rangle}|\bigr)\bigl( \sum_{k=0}^{\infty}
|e^{-k\langle h,\delta\rangle}|\bigr)^N.$$
The first sum is finite; for the second sum one has
$$\sum_{k=0}^{\infty}
|e^{-k\langle h,\delta\rangle}|=\sum_{k=0}^{\infty}
\bigl(e^{-\Ree\langle h,\delta\rangle}\bigr)^{k}<\infty\
\text{ if } \Ree\langle h,\delta\rangle>0.$$ 

\subsubsection{}
\begin{rem}{remRr}
This argument
implies the following: $\ch_{M(\lambda)}$ converges
in the domain $Y_{>}:=\{h\in\fh|\ \Ree\alpha_i(h)>0,\ i=0,1,\ldots,n\}$
to a holomorphic function. The same holds for
the character $\ch_{V(\lambda)}(h)$ since it 
 is majorized
by $\ch_{M(\lambda)}(\Ree(h))$ (where $\Ree(h)\in\fh_{\mathbb{R}}$
is such that $\alpha(\Ree(h))=\Ree\alpha(h)$ for any $\alpha\in\Delta$);
for $h\in Y_{>}$ one has $\Ree(h)\in Y_{>}$. Since
the summands in $\ch_{M(\lambda)}(\Ree(h))$
are positive real numbers $\ch_{V(\lambda)}(h)$ is holomorphic
in $Y_{>}$ by the Weierstrass criterion.
\end{rem}

\subsection{}
Since $R(h)$ is holomorphic in $Y$, 
$\ch_{M(\lambda)}(h)=R^{-1}(h)e^{\langle h,\lambda\rangle}$
is meromorphic in $Y$ for any $\lambda$. 

\subsubsection{}
Consider the case when $(\Lambda+\rho,\delta)\not\in\mathbb{Q}_{\geq 0}$.
By~\ref{alt} $\ch_{L(\Lambda)}$ is given by~\Thm{thm11} (ii) so
it is a finite sum
of functions $\ch_{M(y.\lambda)}(h)$. Hence $\ch_{L(\Lambda)}(h)$ 
 is meromorphic in $Y$. 

The assumption 
$(\Lambda+\rho,\delta)\not\in\mathbb{Q}_{\geq 0}$
implies that 
$\langle\alpha^{\vee},\Lambda+\rho\rangle\not\in\mathbb{Z}_{>0}$
for $\alpha^{\vee}\in\Delta_j$ with $j>>0$ (see~\ref{Xk}
for notation). Then~\cite{KK}, $M(\Lambda)$ has  finite length.
 Therefore
$\ch_{V(\Lambda)}$ is a finite linear combination 
of $\ch_{L(w.\Lambda)}(h)$. Clearly, 
$(w.\Lambda+\rho,\delta)=(\Lambda+\rho,\delta)$
and so, by above,
$\ch_{L(w.\Lambda)}(h)$ is  meromorphic in $Y$ for any $w$.
Hence $\ch_{V(\Lambda)}$  is meromorphic in $Y$.

\subsubsection{}
Therefore it remains to consider the case
when $(\Lambda+\rho,\delta)\in\mathbb{Q}_{>0}$. 

\subsection{}
Fix $\Lambda$ such that $(\Lambda+\rho,\delta)\in\mathbb{Q}_{>0}$.
Take $\lambda\in\mathcal{C}^+$ and $x\in W(\lambda)$ such that 
$\Lambda=x.\lambda$.

In this case
$\ch_{L(\Lambda)}$ is given by~\Thm{thm11} (i): 
$$
\ch_{L(\Lambda)}(h)=\ch_{L(x.\lambda)}(h)=
R^{-1}(h)\sum_{z\in W(\lambda):\ z\geq_{\lambda} x}
(-1)^{l_{\lambda}(z)-l_{\lambda}(x)}Q^{\lambda}_{z,y}(1)
e^{\langle h,z.\lambda\rangle}
$$
for some $\lambda\in\mathcal{C}^+,\ x\in W(\lambda)$ (see~\ref{notKT}
for notation).

All simple subquotients of $M(\Lambda)=M(x.\lambda)$ are of the form
$L(y.\lambda)$ for some $y\geq_{\lambda} x$. By~\Lem{sl}, the condition 
$y\geq_{\lambda} x$ ensures that $y$
is the maximal element of $yW_0(\lambda)$. Combining~\Thm{thm11}
and the definition of $Q_{x,y}$ one sees that
the multiplicity of $L(y.\lambda)$ in $M(x.\lambda)$ is 
$P_{x,y}^{\lambda}(1)$. Since $V(\Lambda)=V(x.\lambda)$ 
is a quotient of $M(x.\lambda)$, and the multiplicity of 
$L(y.\lambda)$ in $M(x.\lambda)$ is 
$P_{x,y}^{\lambda}(1)$~\cite{KT},
the multiplicity of $L(y.\lambda)$ in $V(x.\lambda)$ is 
not greater than $P_{x,y}^{\lambda}(1)$, that is
$$\ch_{V(x.\lambda)}(h)=\sum_{y\in W(\lambda):\ y\geq_{\lambda} x}
a_y \ch_{L(y.\lambda)}(h),\ \ \text{ for some } a_y\in\mathbb{Z}_{\geq 0},\ 
\ a_y\leq P_{x,y}^{\lambda}(1).$$

Hence
\begin{equation}\label{kacav}\begin{array}{l}
|R(h)\ch_{V(x.\lambda)}(h)|\leq\sum_{z\in W(\lambda):\ z\geq_{\lambda} x}
b_z e^{\Ree\langle h,z.\lambda\rangle}, \ \text{ where }\\
b_z=\sum_{y\in W(\lambda):\ x\leq_{\lambda} y\leq_{\lambda} z} 
 |P_{x,y}^{\lambda}(1)Q_{y,z}^{\lambda}(1)|.
\end{array}
\end{equation}

From now on we fix $\lambda\in\mathcal{C}^+$ such that
$(\lambda+\rho,\delta)\in\mathbb{Q}_{>0}$
and fix $x\in W(\lambda)$.

\subsubsection{}\label{ab}
Retain notation of~\ref{cl}. For $a,b>0$ set
$$Y_{a,b}:=\{h\in\fh:  
\Ree\langle h,\delta\rangle>a,\  |\langle h,\alpha\rangle|<b\ 
\text{ for any }\ \alpha\in\Delta_{cl}\}.$$

\subsubsection{}
In view of~(\ref{kacav}), by the Weierstrass criterion, it remains
to verify that  the sum
\begin{equation}\label{fdef}
\sum_{l=0}^{\infty}\bigl(\sum_{z\in W(\lambda):\ z\geq_{\lambda}x,\ l(z)=l}
\sum_{y\in W(\lambda):\ x\leq_{\lambda} y\leq_{\lambda} z} 
 |P_{x,y}^{\lambda}(1) Q^{\lambda}_{x,y}(1)| \bigr)
e^{\Ree\langle h,z.\lambda-\lambda\rangle}
\end{equation}
converges to a uniformly bounded function of $h$
in the domain $Y_{a,b}$, for any $a,b>0$.

\subsection{}
\begin{prop}{propim}
There exists $C>0$ such that
for all but finitely many $y\in W(\lambda)$ one has
$$\Ree\langle h,\lambda-y.\lambda\rangle > Cl_{\lambda}(y)^2
\ \ \ \text{ for any }\ 
h\in Y_{a,b}.$$
\end{prop}
We prove this proposition in~\ref{pims}-\ref{pimi} below.

\subsubsection{}\label{pims}
Retain notation of~\ref{notKT} and
set $\Delta^{\pm}(\lambda):=\Delta(\lambda)\cap\Delta^{\pm}$,
$\Pi_0(\lambda):=\{\alpha\in\Pi(\lambda)|\  
\langle\alpha^{\vee},\lambda+\rho\rangle=0\}$.
Recall that $W_0(\lambda)$ is a finite group
generated by $\{s_{\alpha}| \alpha\in \Pi_0(\lambda)\}$;
let $L$ be the length of the longest elements of $W_0(\lambda)$.

\subsubsection{}
Fix $y\in W(\lambda)$ and write its reduced
decomposition in the Coxeter group $W(\lambda)$ in the form
$y=y_1s_{i_1}y_2s_{i_2}\ldots y_ms_{i_m}y_{m+1}$, where
$y_1,\ldots, y_{m+1}\in W_0(\lambda)$ and each $s_{i_j}$
is the reflection with respect to 
$\alpha_{i_j}\in \Pi(\lambda)\setminus\Pi_0(\lambda)$.
By above, $l_{\lambda}(y_j)\leq L$, so 
\begin{equation}\label{lam}
m> \frac{l_{\lambda}(y)}{L+1}-1. \end{equation}

\subsubsection{}\label{disco}
Let $X$ be the Dynkin
diagram of $\Pi(\lambda)$ constructed as in~\ref{secXi}.
By~\Prop{propXi}, $X$ is a disjoint union of Dynkin
diagrams of finite and affine types; denote by $\Pi'$ the set of 
simple roots for the diagram $X$ and by $\Delta'$ the set of roots
for $X$ (the elements of $\Pi'$ are linearly independent whereas
the elements of $\Pi(\lambda)$ may be linearly dependent, 
see~\Prop{propXi} (ii)). Extend the natural identification
of $\Pi'$ and $\Pi(\lambda)$ to the linear map
$\iota: \mathbb{R}\Pi'\to\mathbb{R}\Pi(\lambda)$.
Retain notation of~\ref{luara}.
For $w\in W(\lambda)$ set $S'(w):=\Delta'_+\cap w^{-1}(\Delta'_-)$.
By~\ref{luara}, the elements of $S'(w)$ are pairwise distinct.

From~\Lem{lara} we obtain
\begin{equation}\label{lay}
\lambda-y.\lambda=(\lambda+\rho)-y(\lambda+\rho)=
\sum_{j=1}^m\langle\alpha^{\vee}_{i_j},\lambda+\rho\rangle\iota(\beta_j'),
\end{equation}
where $\{\beta'_j\}_{j=1}^m$ is a subset of 
$S'(y^{-1})\subset \Delta_{+}^{re}$. We claim that
the elements $\iota(\beta'_j)$ are pairwise distinct. Indeed,
take $\beta'_i,\beta'_j\in S'(y^{-1})$ such that
$\beta'_i-\beta'_j\in\Ker\iota$. Each element of $\Delta'$
lies in the root system of a certain connected component of $X$.
Let $\beta'_i\in\Delta(X_s),\ \beta'_j\in\Delta(X_t)$.
By~\Prop{propXi} (i), $\Ker\iota\cap \Delta(X_s)=0$ so
$\beta'_i-\beta'_j\not\in\Ker\iota$ if $s=t$. 
By~\Prop{propXi} (ii) for $s\not=t$
the assumption $\beta'_i-\beta'_j\in\Ker\iota$ gives 
$\iota(\beta'_i)\in\mathbb{C}\delta$. This contradicts  the fact that
$\beta'_i$ is a real root.

By above, $\beta_j:=\iota(\beta_j)$ ($j=1,\ldots,m$) 
are pairwise distinct elements of $\Delta_{+}^{re}$; 
set $R(y):=\{\beta_j\}_{j=1}^m$.

\subsubsection{}\label{aa}
Fix $h\in Y_{a,b}$.
Take $D\in\fh^*$ such that $\langle D,\Delta_{cl}\rangle=0,\ \ 
\langle D,\delta\rangle=1$. Write $h=a'D+h'$, where 
$h'$ lies in the span of $\Delta$ and notice that $\Ree a'>a$.
One has
$\lambda-y.\lambda=\langle D,\lambda-y.\lambda\rangle\delta+
cl(\lambda-y.\lambda)$ and so
$$\langle h,\lambda-y.\lambda\rangle=a'
\langle D,\lambda-y.\lambda\rangle+\langle h',\lambda-y.\lambda\rangle
=a'\langle D,\lambda-y.\lambda\rangle+\langle h,
cl(\lambda-y.\lambda)\rangle.$$

\subsubsection{}\label{bb}
Recall that 
$\langle\alpha^{\vee}_{i_j},\lambda+\rho\rangle\in\mathbb{Z}_{>0}$ for
$j=1,\ldots,m$ and that 
$\langle D,\gamma\rangle\in\mathbb{Z}_{\geq 0}$ for any $\gamma\in\Delta_+$.
 From~(\ref{lay}) we get
$$\langle D,\lambda-y.\lambda\rangle\geq 
\sum_{j=1}^m\langle D,\beta_j\rangle.$$
Setting
$b':=\max_{\alpha\in\Pi(\lambda)}\langle\alpha^{\vee},\lambda+\rho\rangle$,
we obtain from~(\ref{lay})
$$|\langle h,cl(\lambda-y.\lambda)\rangle|\leq 
\sum_{j=1}^m|\langle\alpha^{\vee}_{i_j},\lambda+\rho\rangle|
|\langle h,cl(\beta_j)|< mbb'\ \text{ for } h\in Y_{a,b}.$$

\subsubsection{}
Retain notation of~\ref{disco}.
 By~\ref{Xk}, the cardinality of 
$X_k=\{\beta\in\Delta_+|\ \langle D,\beta\rangle< k\}$ 
is not greater than $Nk$ so
$$|\{\beta\in R(y)|\ \langle D,\beta\rangle \geq k\}|=
|R(y)\setminus X_k|>m-Nk.$$
Set $m':=[m/N]$. Then
$$\begin{array}{rl} \sum_{j=1}^m\langle D,\beta_j\rangle&=
\sum_{k=1}^{\infty}
|\{\beta\in R(y)|\ \langle D,\beta\rangle \geq k\}|>\sum_{k=1}^{m'}
|\{\beta\in S_{\lambda}(w)|\ \beta\rangle \geq k\}|\\
& >\sum_{k=1}^{m'} (Nm'-Nk)=Nm'(m'-1)/2>
\frac{N}{2}(\frac{m}{N}-1)(\frac{m}{N}-2).
\end{array}$$
Hence
\begin{equation}\label{rhow}
\sum_{j=1}^m\langle D,\beta_j\rangle>\frac{m^2}{8N} \text{ if }
m>4N.\end{equation}

\subsubsection{}\label{pimi}
By~(\ref{lam}), $m\to\infty$ if $l_{\lambda}(y)\to\infty$. 
Now~\ref{aa},\ref{bb} and~(\ref{rhow}) give
$$\Ree\langle h,\lambda-y.\lambda\rangle\geq a\frac{m^2}{8N}-mbb' >
a\frac{m^2}{16N}
\text{ for } l_{\lambda}(y)>>0.$$
Combining with~(\ref{lam}) we obtain for $M:=2^7L^2N$
$$\Ree\langle h,\lambda-y.\lambda\rangle\geq a\frac{l_{\lambda}(y)^2}{M}
\text{ for } l_{\lambda}(y)>>0.$$
This completes the proof of~\Prop{propim}. \qed

\subsection{}\label{recall}
Recall that the Coxeter group
of affine type is a semidirect product of a finite Weyl group  and
a free abelian group of  finite rank, so it has  polynomial growth.
By~\ref{Waf}, $W(\lambda)$ is a direct
product of Coxeter groups of finite and
affine types. In particular, $W(\lambda)$ is of  polynomial growth, i.e.
\begin{equation}\label{bubu}
\exists C'>0, k\geq 0\ \text{ such that }\  
\ |\{w\in W(\lambda)| l_{\lambda}(w)\leq j\}|<C'j^k\ \text{ for all } j\geq 1,
\end{equation}
and the estimate in~\ref{Qestim} is valid for $Q^{\lambda}_{x,y}$.

\subsubsection{}
Combining~(\ref{bubu}), (\ref{Pestim}) and~\Lem{Qestim} we obtain
that there exist $C_1>C',N>k$ such that for any $z\in W(\lambda)$ one has
$$\sum_{y\in W(\lambda):\ x\leq_{\lambda} y\leq_{\lambda} z} 
|P^{\lambda}_{x,y}(1)Q^{\lambda}_{y,z}(1)|<
(C_1 l_{\lambda}(z)^N)^{l_{\lambda}(z)}.$$

Combining this inequality with~\Prop{propim} and 
using~(\ref{bubu}) again, we get for all but finitely many $l>0$ 
$$\sum_{z\in W(\lambda): z\geq_{\lambda}x,l(z)=l}\
\sum_{y\in W(\lambda):x\leq_{\lambda} y\leq_{\lambda} z} 
 |P_{x,y}^{\lambda}(1) Q^{\lambda}_{x,y}(1)| 
e^{\Ree\langle h,z.\lambda-\lambda\rangle}<
(C_1 l^{N})^{l+1} e^{-Cl^2}$$
   for any $h\in Y_{a,b}$. Since the series
$\sum_{l=0}^{\infty} (C_1 l^{N'})^{l+1} e^{-Cl^2}$ converges,
the sum~(\ref{fdef}) is uniformly bounded for $h\in Y_{a,b}$.
This completes the proof of~\Thm{propmain}.

\subsection{}
\begin{rem}{}
We do not know whether the characters of highest weight modules 
over arbitrary  Kac-Moody algebra $\fg$ are meromorphic
functions in the interior of the complexified Tits cone $Y$ (see~\cite{Kbook},
10.6 for its definition). However, in the rank 2 case it is true, provided
that $\Lambda+\rho$ lies in the dual complexified Tits cone.
Indeed, the denominator of $\ch_{V(\lambda)}$ equals
to $\sum_{w\in W} (-1)^{l(w)} e^{w\rho-\rho}$ by the denominator identity,
which converges to a holomorphic function in $Y$~[K, 10.6].
The numerator in this case is always of the form 
$\sum_w (-1)^{l(w)} e^{w(\Lambda+\rho)-\rho}$, 
where $w$ runs over a subset of $W$,
since in the rank 2 case $P_{x,y}=1$ for all $x\leq y$. 
This series converges to a holomorphic function in $Y$ as well.
\end{rem}

\section{The poles of $\ch_{V(\lambda)}(h)$}
In this section $\lambda\in\fh^*$ is non-critical, i.e.
$(\lambda+\rho,\delta)\not=0$.

By~\Thm{propmain}, the meromorphic function 
$\ch_{V(\lambda)}(h)$ in $Y$ has at most simple
poles at the hyperplanes $\alpha=0$, where
$\alpha\in\Delta^{re}_+$, and no other singularities.
The collection $S_{V(\lambda)}\subset \Delta^{re}_{+}$ of $\alpha$'s,
for which $\ch_{V(\lambda)}(h)$ does have a singularity, is an interesting
invariant of $V(\lambda)$. This kind of
invariant has been studied in the finite-dimensional case
under the name Borho-Jantzen-Duflo 
$\tau$-invariant (see~\cite{BJ},\cite{BJo},\cite{V} and references there).

If $V(\lambda)$ is integrable then $\ch_{V(\lambda)}(h)$ is holomorphic
in $Y$ (see~\cite{Kbook}, 11.10).
In this section we will show that the converse is true: if $\lambda$ is
non-critical and $\ch_{V(\lambda)}(h)$ is holomorphic in $Y$, 
then $V(\lambda)$ is integrable. We also show that if $\lambda$ is
non-critical and $S_{L(\lambda)}=\Delta^{re}_{+}$, then
$L(\lambda)$ is  a Verma module.

\subsection{}
\begin{prop}{propxla}
Fix $\lambda\in\mathcal{C}$. Let $\alpha\in\Pi(\lambda)$ 
be such that $M(s_{\alpha}.\lambda)$ is a submodule of 
$M(\lambda)$. If $V(\lambda)$ is a subquotient of 
$M(\lambda)/M(s_{\alpha}.\lambda)$
then $\alpha\not\in S_{V(\lambda)}$.
\end{prop}
\begin{proof}
Clearly, we may assume that $M(s_{\alpha}.\lambda)\not
=M(\lambda)$. One has
$$\ch_{M(\lambda)/M(s_{\alpha}.\lambda)}=e^{\lambda}\frac{1-e^{-k\alpha}}
{\prod_{\beta\in\Delta_+}(1-e^{-\beta})}=e^{\lambda}(1+e^{-\alpha}+\ldots+
e^{-(k-1)\alpha})\prod_{\beta\in\Delta_+\setminus\{\alpha\}}
(1-e^{-\beta})^{-1}
$$
for some $k>0$.
We claim that $\ch_{M(\lambda)/M(s_{\alpha}.\lambda)}(h)$ absolutely
converges in the domain
$$Y_{>;\alpha}:=
\{h\in\fh|\ \forall \beta\in\Delta_+\setminus\{\alpha\}\ \ 
\Ree\beta(h)>0\}.$$
Indeed, in $Y_{>;\alpha}$ the absolute convergence of the infinite product
$\prod_{\beta\in\Delta_+\setminus\{\alpha\}}
(1-e^{-\beta(h)})^{-1}$ is equivalent to the absolute convergence of 
the infinite sum $\sum_{\beta\in\Delta_+\setminus\{\alpha\}}
e^{-\beta(h)}$, which converges absolutely in 
 $Y_{>;\alpha}$ by the argument of~\ref{Rr}.

Since $\ch_{V(\lambda)}(h)$ is majorized by 
$\ch_{M(\lambda)/M(s_{\alpha}.\lambda)}(h)$, $\ch_{V(\lambda)}(h)$
is holomorphic in $Y_{>;\alpha}$.

Consider the case when $\alpha\in\Pi$.
Consider $\fh$ as a vector space over $\mathbb{R}$.
The domain $Y$ is bounded by the hyperplanes $\Ree\beta(h)=0$, 
$\beta\in \Pi$ so $Y_{>;\alpha}$ strictly contains 
$Y$. Therefore
$Y_{>;\alpha}$ contains an open ball $B$ which meets
the hyperplane $\alpha=0$. Since $\ch_{V(\lambda)}(h)$
is holomorphic in $B$, it does not have a pole at the hyperplane $\alpha=0$
so $\alpha\not\in S_{V(\lambda)}$ as required.

Now consider the case when $\alpha\in\Pi(\lambda)\setminus\Pi$.
Let $X$ be the Dynkin diagram of $\Pi(\lambda)$ and let
$\fg_{\lambda}$ be the Kac-Moody algebra corresponding to $X$.
By~\Prop{propXi} (iii), $\fg_{\lambda}$ is the direct product of 
simple finite dimensional and affine Lie algebras.
Let $\fh_{\lambda}$ be the Cartan subalgebra of $\fg_{\lambda}$
and $\Pi'\subset \fh_{\lambda}^*$ be the set of simple roots.
We will reduce the assertion that $\ch_{V(\lambda)}(h)$
does not have a pole at the hyperplane $\alpha=0$
to a similar assertion for a highest weight module over $\fg_{\lambda}$.

Extend the natural identification of $\Pi'$ and $\Pi(\lambda)$
to the linear map $\iota:\mathbb{C}\Pi'\to \mathbb{C}\Pi(\lambda)$.
Let $\rho_{\lambda}$ be the standard Weyl vector for $\Delta(\lambda)_+$;
introduce $*$-action of $W(\lambda)$ on $\fh^*_{\lambda}$ and on $\fh^*$ by
the formulas
$$w*\mu=w(\mu+\rho_{\lambda})-\rho_{\lambda},\ \ \ 
w*\mu=w(\mu+\iota(\rho_{\lambda}))-\iota(\rho_{\lambda}).$$  
For $\nu\in\fh^*_{\lambda}$ denote by $M'(\nu)$ (resp., by $L'(\nu)$)  the 
Verma (resp., simple) $\fg_{\lambda}$-module of highest weight $\nu$.
Let $\pi:\fh^*\to \mathbb{C}\Pi(\lambda)=\im\iota$ be the 
orthogonal projection (i.e., $\ker\pi=\{\mu\in\fh^*|\ (\mu,
\mathbb{C}\Pi(\lambda))=0\}$). Set 
$\nu:=\iota^{-1}\pi(\lambda+\rho)-\rho_{\lambda}$.
Then $\lambda+\rho-\iota(\nu+\rho_{\lambda})$ lies in $\ker\pi$.
Since $W(\lambda)$ stabilizes the elements of $\ker\pi$ one has
$$\lambda-w.\lambda=\iota(\nu-w*\nu) \ \text{ for any } w\in W(\lambda).$$
In particular, writing $\lambda-s_{\alpha}.\lambda=k\alpha$
we obtain $\nu-s_{\alpha}*\nu=k\alpha$.
By the assumption, $M(s_{\alpha}.\lambda)$ is a proper submodule
of $M(\lambda)$ so $k$ is a positive
integer. Therefore $M'(s_{\alpha}*\nu)$ is a submodule of $M'(\nu)$.

Observe that for any $w\in W(\lambda)$ one has
$$e^{-\lambda}\prod_{\beta\in\Delta_+\setminus\Delta(\lambda)}(1-e^{-\beta})
\ch_{M(w.\lambda)}=\iota\bigl(e^{-\nu}\ch_{M'(w.\nu)}\bigr),$$
where $\iota(e^{\xi}):=e^{\iota(\xi)}$.
From \Thm{thm11} (\cite{KT}) we obtain for any $x\in W(\lambda)$
$$e^{-\lambda}\prod_{\beta\in\Delta_+\setminus\Delta(\lambda)}(1-e^{-\beta})
\ch_{L(x.\lambda)}=\iota\bigl(e^{-\nu}\ch_{L'(x.\nu)}\bigr).$$

Recall that if $L(\lambda')$ is a subquotient of $M(\lambda)$ then
$\lambda'\in W(\lambda).\lambda$. Let $a_w,\ b_w,\ c_w$ be the multiplicities
 of $L(w.\lambda)$ in $V(\lambda)$, in $M(s_{\alpha}.\lambda)$
and in $M(\lambda)$ respectively. Clearly, 
$a_w\leq c_w-b_w$. One has $\ch_{V(\lambda)}=\sum_{w\in W(\lambda)}
a_w \ch_{L(w.\lambda)}$ so
\begin{equation}\label{ask}
e^{-\lambda}\prod_{\beta\in\Delta_+\setminus\Delta(\lambda)}(1-e^{-\beta})
\ch_{V(\lambda)}=\sum_{w\in W(\lambda)}
a_w \iota\bigl(e^{-\nu}\ch_{L'(w.\nu)}\bigr).
\end{equation}
For $h\in Y$ set 
$$f(h):=e^{-\lambda(h)}\prod_{\beta\in\Delta_+\setminus\Delta(\lambda)}
(1-e^{-\beta(h)})\ch_{V(\lambda)}(h).$$
Recall that $f(h)$ is a meromorphic function on $Y$. 

Denote by $(\mathbb{C}\Pi_{\lambda})^{\perp}$ the orthogonal to
$\mathbb{C}\Pi_{\lambda}$, i.e.
$$(\mathbb{C}\Pi_{\lambda})^{\perp}=\{h\in\fh|\ \beta(h)=0\ \text{ for any }
\beta\in\Pi(\lambda)\},$$
and by $(\mathbb{C}\Pi')^{\perp}$ the orthogonal to $\mathbb{C}\Pi'$, i.e.
$$(\mathbb{C}\Pi')^{\perp}=\{h\in\fh_{\lambda}|\ \beta(h)=0\ \text{ for any }
\beta\in\Pi'\}.$$
Any $\beta\in\Pi(\lambda)$ defines a functional on
$\fh/(\mathbb{C}\Pi_{\lambda})^{\perp}$ and any $\beta\in \Pi'$
defines a functional on $\fh_{\lambda}/(\mathbb{C}\Pi')^{\perp}$.

Since $\im\iota=\mathbb{C}\Pi_{\lambda}$ from~(\ref{ask})
one sees that $f(h)=f(h+h')$ if 
$h,h+h'\in Y$ and $h'\in (\mathbb{C}\Pi_{\lambda})^{\perp}$.
Denote by $F$ the corresponding function on 
$Y/(\mathbb{C}\Pi')^{\perp}$: 
$$F(h+(\mathbb{C}\Pi')^{\perp}):=f(h)
=e^{-\lambda(h)}\prod_{\beta\in\Delta_+\setminus\Delta(\lambda)}
(1-e^{-\beta(h)})\ch_{V(\lambda)}(h).$$
In order to show that $f(h)$ does not have a pole at the hyperplane
$\alpha=0$, it is enough to show that $F(h)$ 
does not have a pole at the hyperplane $\alpha=0$.

Recall that $\ch_{L'(w.\nu)}(h)$ is a  meromorphic function on 
$$Y':=\{h\in \fh_{\lambda}| \Ree \delta_i(h)>0\ \text{ for } i=1,\ldots,k\},$$
where $\delta_1,\ldots,\delta_k$ are the minimal imaginary roots
of the affine components of $\fg_{\lambda}$. Observe that
$e^{-\nu}\ch_{L'(w.\nu)}$ is a linear combination of
$e^{\mu}$, where $\mu\in\mathbb{Z}\Pi'$. Therefore
$e^{-\nu}\ch_{L'(w.\nu)}(h)=e^{-\nu}\ch_{L'(w.\nu)}(h+h')$
if $h,h+h'\in Y'$ and $h'\in (\mathbb{C}\Pi')^{\perp}$.
Denote by $F_w$ the function on $Y'/(\mathbb{C}\Pi')^{\perp}$
defined by 
$$F_w(h+(\mathbb{C}\Pi')^{\perp}):=e^{-\nu}\ch_{L'(w.\nu)}(h).$$

Identify $\fh/(\mathbb{C}\Pi_{\lambda})^{\perp}$ with
$(\mathbb{C}\Pi_{\lambda})^*$ and $\fh_{\lambda}/(\mathbb{C}\Pi')^{\perp}$
with $(\mathbb{C}\Pi')^*$. 
Denote by $\iota^*$ the map $\fh/(\mathbb{C}\Pi_{\lambda})^{\perp}\to
\fh_{\lambda}/(\mathbb{C}\Pi')^{\perp}$ which is conjugate to
$\iota:\mathbb{C}\Pi'\to \mathbb{C}\Pi_{\lambda}$. 
By~(\ref{ask}) one has
$$F(h)=\sum_{w\in W(\lambda)} a_w F_w(\iota^*(h)).$$
In order to show that $F(h)$ does not have a pole
at $\alpha=0$, we will show that the series in the right-hand side
absolutely converges in the domain
$$Y(\lambda)_{>;\alpha}:=
\{h\in \fh/(\mathbb{C}\Pi_{\lambda})^{\perp}|\ \Ree\beta(h)>0\ \text{ for any }
\beta\in\Delta(\lambda)_+\setminus\{\alpha\}\}.$$
Indeed, by~\Thm{thm11}, $b_w$ is equal to the multiplicity
of $L'(w*\nu)$ in $M'(s_{\alpha}*\nu)$ and $c_w$ is equal to the multiplicity
of $L'(w*\nu)$ in $M'(\nu)$. Therefore the series $\sum_{w\in W(\lambda)}
a_w \ch_{L'(w.\nu)}(h)$ is majorized
by the series $\ch_{M'(\nu)/M'(s_{\alpha}*\nu)}(h)$. 
Recall that $\alpha\in\Pi(\lambda)$ so, by above, the series
$\ch_{M'(\nu)/M'(s_{\alpha}*\nu)}(h)$ absolutely converges in the domain
$$Y'_{>;\alpha}:=\{h\in\fh_{\lambda}|\ \Ree\beta(h)>0\ \text{ for any }
\beta\in\Delta(\lambda)_+\setminus\{\alpha\}\}.$$ 
Clearly, $\iota^*(Y(\lambda)_{>;\alpha})=Y'_{>;\alpha}/
(\mathbb{C}\Pi')^{\perp}$. The claim follows.
Hence  $\ch_{V(\lambda)}(h)$ does not have a pole at $\alpha=0$.
\end{proof}

\subsubsection{}
Retain notation of~\ref{C+-}.

\begin{cor}{corxla}
(i) Let $\lambda\in\mathcal{C}^+$ and let $x\in W(\lambda)$ be the longest
element in a coset $xW_0(\lambda)$. If
$\alpha\in\Pi(\lambda)$ is such $s_{\alpha}x>_{\lambda} x$,
then $\alpha\not\in S_{L(x.\lambda)}$.

(ii) Let $\lambda\in\mathcal{C}^-$ and let $x\in W(\lambda)$ be the shortest
element in a coset $xW_0(\lambda)$. If
$\alpha\in\Pi(\lambda)$ is such $s_{\alpha}x<_{\lambda} x$,
then $\alpha\not\in S_{L(x.\lambda)}$.

(iii) If $\Lambda$ is non-critical and $S_{L(\Lambda)}=\Delta_+$,
then $L(\Lambda)$ is a Verma module.
\end{cor}
\begin{proof}
 The inclusion  $M(s_{\alpha}x.\lambda)\subset M(x.\lambda)$ 
is equivalent to the inequality $s_{\alpha}x\geq_{\lambda} x$ 
(resp., $s_{\alpha}x\leq_{\lambda} x$) for $\lambda\in\mathcal{C}^+$
(resp., for $\lambda\in\mathcal{C}^-$). The inclusion is proper
due to the assumption that $x\in W(\lambda)$ is the longest 
(resp., the shortest)
element in a coset $xW_0(\lambda)$.

For (iii) write $\Lambda=x.\lambda$
for $x\in W(\Lambda),\ \lambda\in \mathcal{C}^+\cup \mathcal{C}^-$,
where $x$ is the longest (resp., the shortest)
element in a coset $xW_0(\lambda)$ if $\lambda\in \mathcal{C}^+$
(resp.,  $\lambda\in \mathcal{C}^-$). Combining the assumption 
$S_{L(\Lambda)}=\Delta_+$ and (i), (ii) we conclude that
$\lambda\in\mathcal{C}^-$ and $x=e$. Then $M(\Lambda)=M(\lambda)$
is simple, so $L(\Lambda)=M(\Lambda)$ as required.
\end{proof}

\subsubsection{}
\begin{rem}{}
It is not true that $S_{V(\Lambda)}=\Delta_+$ implies that $V(\lambda)$
is a Verma module. For example, consider $\fg=\hat{\fsl}(2)$ and 
a highest weight module $V(0):=M(0)/M(s_0s_1.0)=M(0)/M(-3\alpha_0-\alpha_1)$.
One has
$$\ch_{V(0)}=\frac{1-e^{-3\alpha_0-\alpha_1}}{(1-e^{-\alpha_1})
\prod_{k=0}^{\infty}
(1-e^{-k\alpha_0-(k-1)\alpha_1})(1-e^{-k\alpha_0-k\alpha_1})
(1-e^{-k\alpha_0-(k+1)\alpha_1})},$$
hence $S_{V(0)}=\Delta_+$.
\end{rem}

\subsection{}
\begin{prop}{}
Let $V(\Lambda)$ be a highest weight module with non-critical highest weight
$\Lambda$ over an affine Lie algebra. Then the character
$\ch_{V(\Lambda)}(h)$ is holomorphic in
$Y$ if and only if the module $V(\Lambda)$ is integrable
(equivalently, if and only if $V(\Lambda)=L(\Lambda)$ and $\Lambda\in P^+$).
\end{prop}
\begin{proof}
Recall that a highest weight module $V(\Lambda)$ is integrable  
if and only if it is simple and $\Lambda\in P^+$, where
$P^+=\{\lambda\in\fh^*|\ \langle \lambda+\rho,\alpha\rangle\in\mathbb{Z}_{>0}
\text{ for any } \alpha\in\Pi\}$ (see~\cite{Kbook}, Chapter 10). 
Also, by~\cite{Kbook}, 11.10, 
$\ch_{L(\Lambda)}(h)$ is holomorphic in $Y$ if  $L(\Lambda)$ is integrable.

Recall that the series $\ch_{V(\lambda)}(h)$  absolutely converges
in the domain $Y_{>}$ to a holomorphic function
and $\ch_{V(\lambda)}(h)$ is the analytic 
continuation of this function to $Y$. Assume that
this function is holomorphic. We claim that 
the series $\ch_{V(\lambda)}(h)$ absolutely converges in the domain $Y$.
Indeed, write
$$e^{-\lambda(h)}\ch_{V(\lambda)}(h)=\sum_{\nu\in Q^+} m_{\nu}e^{-\nu(h)},
\ \text{
where }\ m_{\nu}=\dim V(\lambda)_{\lambda-\nu}\in\mathbb{Z}_{\geq 0}.$$
Write $\Pi=\{\alpha_i\}_{i=0}^n$, where the simple roots
are arbitrarily numerated, and set $z_i:=e^{-\alpha_i(h)}$
for $i=1,\ldots,n$, $z_0:=e^{-\delta(h)}$. For $\nu\in Q^+$ write
$\nu$ in the form $\nu=k_0\delta+\sum_{i=1}^n k_i\alpha_i$ and 
observe that $k_0\in\mathbb{Z}_{\geq 0},\ \ k_i\in\mathbb{Z}$ for 
$i=1,\ldots,n$; set
$$z^{\nu}:=\prod_{i=0}^n z_i^{k_i}.$$
Rewrite $e^{-\lambda(h)}\ch_{V(\lambda)}(h)$ in new variables
and denote the resulting series by $F(z)$:
$$e^{-\lambda(h)}\ch_{V(\lambda)}(h)=\sum_{\nu\in Q^+} m_{\nu} z^{\nu}=:F(z)$$
By~\Rem{Rr}, the series $F(z)$ converges in the domain 
$$Y_{>}=\{0<|z_i|<1\text{ for } i=1,\ldots,n\ \& \
0<|z_0|^{n_0}\prod_{i=1}^n |z_i|^{n_i}<1\},$$
where $\delta=\sum_{i=0}^n n_i\alpha_i$ so $n_i$ are non-negative integers.
For $i=1,\ldots,n$ fix $z'_i$ such that $|z'_i|<1$. Then the series
$f(z_0):=F(z_0;z'_1,\ldots,z'_n)$ converges in the domain 
$\{0<|z_0|<C\}$ for some $C<1$.  
By above, $f(z_0)$ is a power series so it converges in the open disc
$\{|z_0|<C\}$.
By the assumption, the sum of $F(z)$ can be analytically extended 
to a function which is holomorphic in $Y=\{0<|z_0|<1, z_i\not=0,\text{ for }
i=1,\ldots,n\}$. Therefore the sum of $f(z_0)$ can be analytically extended 
to a function $g(z_0)$
which is holomorphic in the ring $\{0<|z_0|<1\}$.
By above, the  sum of $f(z_0)$ is holomorphic in $\{|z_0|<C\}$.
Thus $g(z_0)$ is holomorphic in the open disc $\{|z_0|<1\}$.
By Cauchy's theorem, this means that the power series $f(z_0)$ converges
to $g(z_0)$ in $\{|z_0|<1\}$;
in particular, $f(z_0)$ absolutely converges in $\{|z_0|<1\}$.
Hence the series $F(z)$ absolutely converges in the domain $Y$.

Now for $i=0,2,\ldots,n$ fix $z'_i\in\mathbb{R}$ such that $0<z'_i<1$. 
Set $f(z_1):=F(z_0';z_1;z_2',\ldots,z'_n)$. By above,
$f(z_1)$ converges for any $z_1\not=0$.
Since $m_{\nu}$ are non-negative integers, the series $f(z_1)$ majorizes
the series $\sum_{k=0}^{\infty} m_{k\alpha_1}z_1^{k}$. Therefore 
$\sum_{k=0}^{\infty} m_{k\alpha_1}x^{k}$ converges for any $x$. 
Since the coefficients
$m_{k\alpha_1}$ are non-negative integers,  we conclude that
$m_{k\alpha_1}=0$ for some $k>0$. This means that
$\dim V(\lambda)_{\lambda-k\alpha_1}=0$. Since $\alpha_1$ is an arbitrary
simple root, we conclude that 
for any $\alpha\in \Pi$ one has 
$\dim V(\lambda)_{\lambda-k\alpha}=0$ for some $k>0$.
By~\cite{Kbook}, Lemma 3.4, this implies
that $V(\lambda)$ is integrable.
\end{proof}


\end{document}